# Two-step MEMS microfabrication via 3D direct laser lithography

Omar Tricinci*[a], Marco Carlotti[a], Andrea Desii[a], Fabian Meder[a], Virgilio Mattoli[a]
[a]Center for Micro-BioRobotics, Istituto Italiano di Tecnologia, Viale Rinaldo Piaggio 34, 56025 Pontedera, Italy

## ABSTRACT

Micro/nano electro-mechanical systems (MEMS/NEMS) are constantly attracting an increasing attention for their relevant technological applications in fields ranging from biology, medicine, ecology, energy to industry. Most of the performances of micro-nanostructured devices rely on both the design and the intrinsic properties of the constituent materials that are processed at such dimensional scale. For this reason, spatial precision, resolution and reproducibility are crucial factors in the micro-fabrication procedure. 3D direct laser lithography (DLL), based on multiphoton absorption, allows to realize outstanding three-dimensional structures with nanoscale features. This technique has recently emerged as a powerful tool for fabricating 3D micro-patterned surfaces for optics, photonics, as well as for bioinspired cell culture scaffold. We propose a method for a two-step fabrication of micro/nanostructured multicomponent systems to be employed as transductors, by means of the integration of 3D DLL and shadowing effects in metal deposition. A z-axis accelerometer is the proof-of-concept for the validation of the proposed transductor. The former is composed of a cantilever patterned with conductive paths which act as a strain gauge. Mechanical stimulation deforms the cantilever and, accordingly, varies its conductive properties. The fabrication of the conductive components is performed using the vacuum evaporation of gold, a traditional microfabrication technique, and exploiting the shadowing effect due to peculiar microstructures on the cantilever.

## 1. INTRODUCTION

In the last decades micro/nano electro-mechanical systems (MEMS/NEMS) have proven themselves as disruptive technology in many fields, ranging from biology, medicine, and ecology to energy and industry[1]. Nowadays, MEMS and NEMS have a considerable impact on different market areas, like automotive, consumer electronics and apparels sectors, and new applications are continuously emerging. Borrowing the same technologies at the base of the integrated circuit industry, these former have introduced a new paradigm for the development of new families of sensors and actuators. In MEMS, both the signal processing unit, the signal acquisition unit, and the actuation unit are embedded in the same micro-device. In fact, micro- and nano-system technologies allow the integration of the different components of a complex device thanks to the proper downscaling at the micrometer and nanometer scale. In this way it's possible to reduce costs while enhancing the performances of the final devices.

Several key points must be considered during the implementation of the MEMS. First of all, the mechanism of the signal transduction can be extremely scale-dependent, since the relative contribution of external physical phenomena can significantly vary according to the dimensional scale[1]. Secondly, also the physical properties of the constituent materials, such as the elastic modulus, can be affected by shape and dimensions. For these reasons the implementation of a MEMS device is strictly focused on the design, that, on the other side is strongly limited by the available microfabrication techniques. Traditional microfabrication techniques employed in MEMS production involve lithography, chemical and physical deposition, dry and wet etching, sputtering, evaporation, thermal treatment, to mention the most common[2]. Therefore, the full potential of MEMS is still largely unexplored, since system development is a time-consuming and costly process, which often requires many consecutive steps, expensive equipment, and long simulation processes.

One may overcome these limitations with novel methodologies for rapid prototyping that can speed up the design and validation processes, and broaden the scope of micro-nanofabrication to structures and functionalities that are not achievable with traditional techniques. In this context, 3D direct laser lithography (DLL), an additive manufacturing technique based on multiphoton absorption, has recently emerged as a powerful tool for the fabrication of 3D micro-structures for optics, photonics and bioinspired applications, also in the biological field[3–10]. Compared to commonly employed 2D (or 2.5D) lithography techniques, 3D microprinting approaches allow unprecedented level of freedom and structure complexity. DLL makes use of a fs-pulsed long-wavelength laser (*e.g.* 780 nm) which is focused, by means of optical elements, in a photosensitive resin able to undergo crosslinking reactions (negative photoresist) or decompose (positive photoresist). By employing resists which are transparent to at the laser wavelength, the laser can be efficiently focused without losses, achieving a high intensity in the focus spot which triggers two (or more)-photon phenomena and the cross-link of the resist. The small volume in which this happen is named "voxel", being the 3D analogue of a 2D pixel. The absorption probability outside of the voxel is small, thus suppressing the propagation of the reaction and resulting in fine resolution (even below 100 nm). By moving around the voxel, one can obtain complex 3D structures that can be easily isolated after developing. In this sense, DLL ensures reproducibility, spatial precision, and resolution at the micro- and nanoscale, two crucial factors in the MEMS/NEMS fabrication process. This technique can excellently replace all the lithographic steps that are required in standard procedures, thus saving time but not at the expenses of the quality of the results. In fact, thanks to this 3D technique it is possible to fabricate almost all the envisioned structures, with high levels of geometrical complexity. While the developing of innovative and smart materials, able to confer the fabricated structures functional properties – such as electrical conductivity, magnetic response, anisotropic actuation – can broaden the application scope of DLL, several solutions can be designed to prepare working devices from readily available commercial materials.

Here we present a novel method for MEMS fabrication consisting in a 3D DLL printing step followed by metal evaporation, is a common microfabrication technique employed for the fabrication of the conductive components. Thanks to presence of 3D-printed self-shadowing structures, the metal evaporation resulted in the formation of independent conductive pathways that could be addressed independently. As proof-of-concept we prepared a transducer. The latter comprised a cantilever patterned with conductive paths which allowed the measurement of the impedance. When the device experiences external acceleration, the cantilever deforms and its conductivity changes significantly. We then characterized properties of the device and its response under different stimuli and frequencies.

## 2. MATERIALS AND METHODS

### 2.1 Design implementation

The design of the cantilever and the relative electrical pad, for the impedance measurements, has been implemented with Inventor. The CAD model is reported in Figure 1a,b. The transducer is composed of a cantilever beam connected to a square pad that is insulated from the rest of the substrate by means of a wall. The cantilever is composed of a double beam with a mass at the end. Onto each branch of the cantilever there is a T-shape wall, as well as along the perimeter of the pad. The T-shape serves to cover part of the structure beneath during the deposition of the metal, exploiting the shadowing effect (Figure 1.c). The same expedient has been implemented for the insulation of one of the bases of the cantilever. In Figure 1.d the predicted flow of the electric current between the measuring electrodes is reported. A similar approach was recently also independently developed by Kim et al. to prepare a voltage-driven micro-actuators.[11]

The transducer was fabricated on a circular glass slide. The sample, after being cleaned with isopropanol (IPA) and rinsed with deionized water, was covered with a nanometric layer of indium tin oxide (ITO), through a sputtering procedure. The presence of ITO is needed for finding the proper interface during the microfabrication.

A 100 µm-thick layer of SU-8 was deposited on the sample via spin coating (1000 rpm for 1 min), followed by two steps of prebake on the hot plate at 65°C for 10 min and 95°C for 30 min. The two-photon polymerization of the photoresist was carried out by means of GT Photonic Professional, Nanoscribe GmbH. The devices were fabricated by exposing the photoresist to a laser beam (Calman laser source) with a center wavelength of 780 nm, by means of an objective (63x, NA 1.4) in oil immersion. The writing speed was 10 mm s$^{-1}$ while the laser power was 13 mW. The sample was then subjected to two steps of soft bake on the hot plate at 65°C for 1 minute and 95°C for 10 min. Finally, it was developed for 20 min in SU-8 Developer (MicroChem Corp) and rinsed with IPA and deionized water.

The sample, with four cantilevers, was than place in the evaporator chamber in order to deposit the conductive layer. A first layer of titanium, with a thickness of 10 nm, was deposited with the purpose to increase the adhesion of the subsequent 40 nm-thick layer of gold.

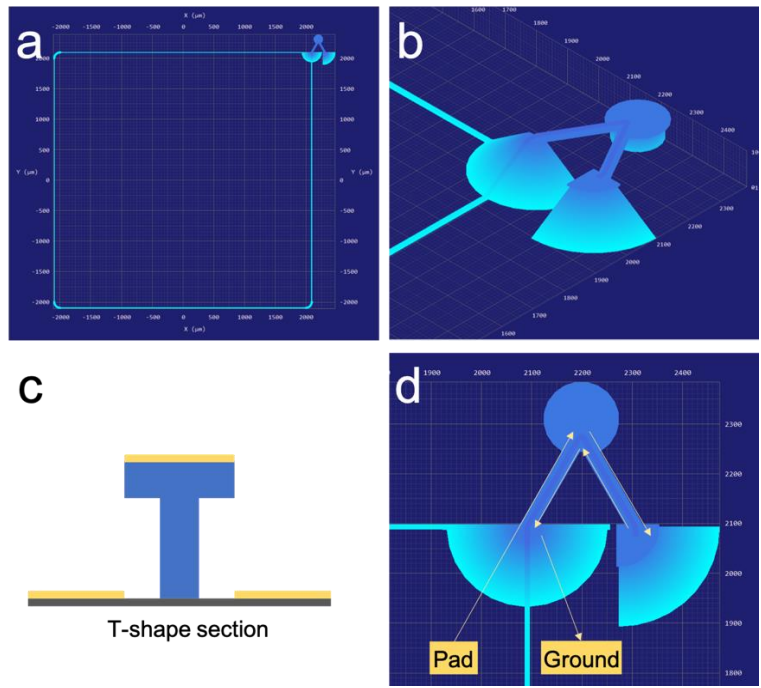

Figure 1. a) Top view of the CAD model of the resonator; b) View of the CAD model of the cantilever beam; c) Cross section of the insulator paths printed on the cantilever and on the wall of the pads; d) Electric paths where changes in conductivity are measured while the cantilever is in oscillation.

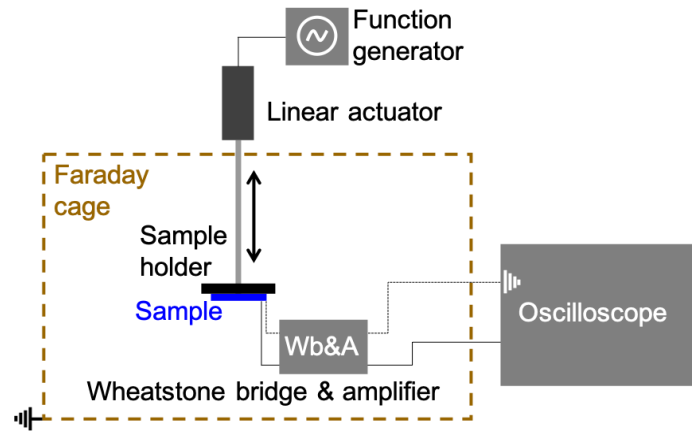

Figure 2. Scheme of the experimental setup for the measurements of the frequency response of the oscillating cantilever.

**2.2 Experimental setup**

Once fabricated, the cantilevers were initially tested in order to check the effectiveness of the electric contact and the absence of short circuits. A resistance in the order of 200 Ω was found for the working devices. Once two copper wires were glued on the pad and on the ground of the cantilever (Figure 1.d), the sample was ready to be mounted in the experimental setup, as described in Figure 2.

Mechanical stimulation as function of frequency and waveform was performed in a self-made apparatus in which a linear actuator (4Ω HiFi full-range driver, diameter 8 cm, model FRS, Visaton, Germany) driven by a monolithic power amplifier (OPA549T, Burr-Brown, USA) and controlled by a function generator (GFG-8217A, GW Instek, Taiwan) was used to apply vertical motion with adjustable waveform (sinusoidal, square), frequency (as indicated), and amplitude

(here ~ 1-2 mm). The sample was attached to a sample holder mounted on the actuator. Voltages were measured by means of an amplified Wheatstone bridge, connected to a passive probe (1:10, 500 MHz, Keysight Technologies, Italy) connected to an oscilloscope (MSO7014A, Agilent Technologies, USA). The setup was placed in a Faraday cage constructed of plain square weave copper mesh using a 0.25 mm copper wire with 1.4 mm spacing (PSY406, Thorlabs, USA). The raw data were analysed using OriginPro 2018 software (b9.5.1.195). The Fast Fourier Transform (FT) was performed using the discrete method in the software.

## 3. RESULTS AND DISCUSSION

Figure 3 shows the results of the microfabrication of the transducer by means of two-photon lithography and single step metal deposition. The results are quite good in terms of resolution and reproducibility. The details of the images in Figure 3 show the successful shadowing effect of the T-shaped walls which isolated the pad from the ground (except for the cantilever path). Remarkably, the inevitable presence of defects at the connection between the printing blocks did not interrupt the conductive metallic path, leaving only the self-shadowing structures to delimitate the designed conductive areas.

Our transducer was conceived to act as accelerometer along the z-direction. Any acceleration experienced by the device along a direction perpendicular to the cantilever plane is expected to bend it and, consequently, inducing a strain on the evaporated 40-nm gold path. In this sense, this latter will act as a strain gauge, responding with a measurable change in conductivity proportional to the deformation, and thus to the acceleration experienced.

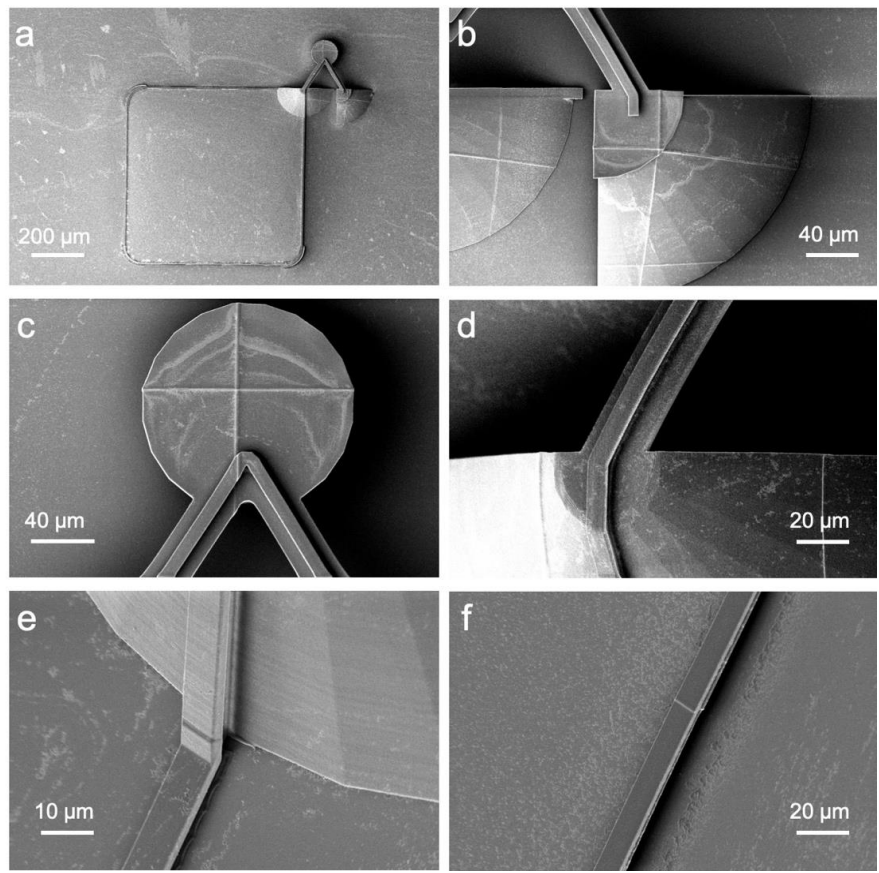

Figure 3. Scanning electron microscope images of the accelerometer: a) Whole device (the pad has been printed in a smaller scale); b) Insulated base of the cantilever; t-shape insulator elements on the beam (d), at the base (e) and in the wall of the pad (f).

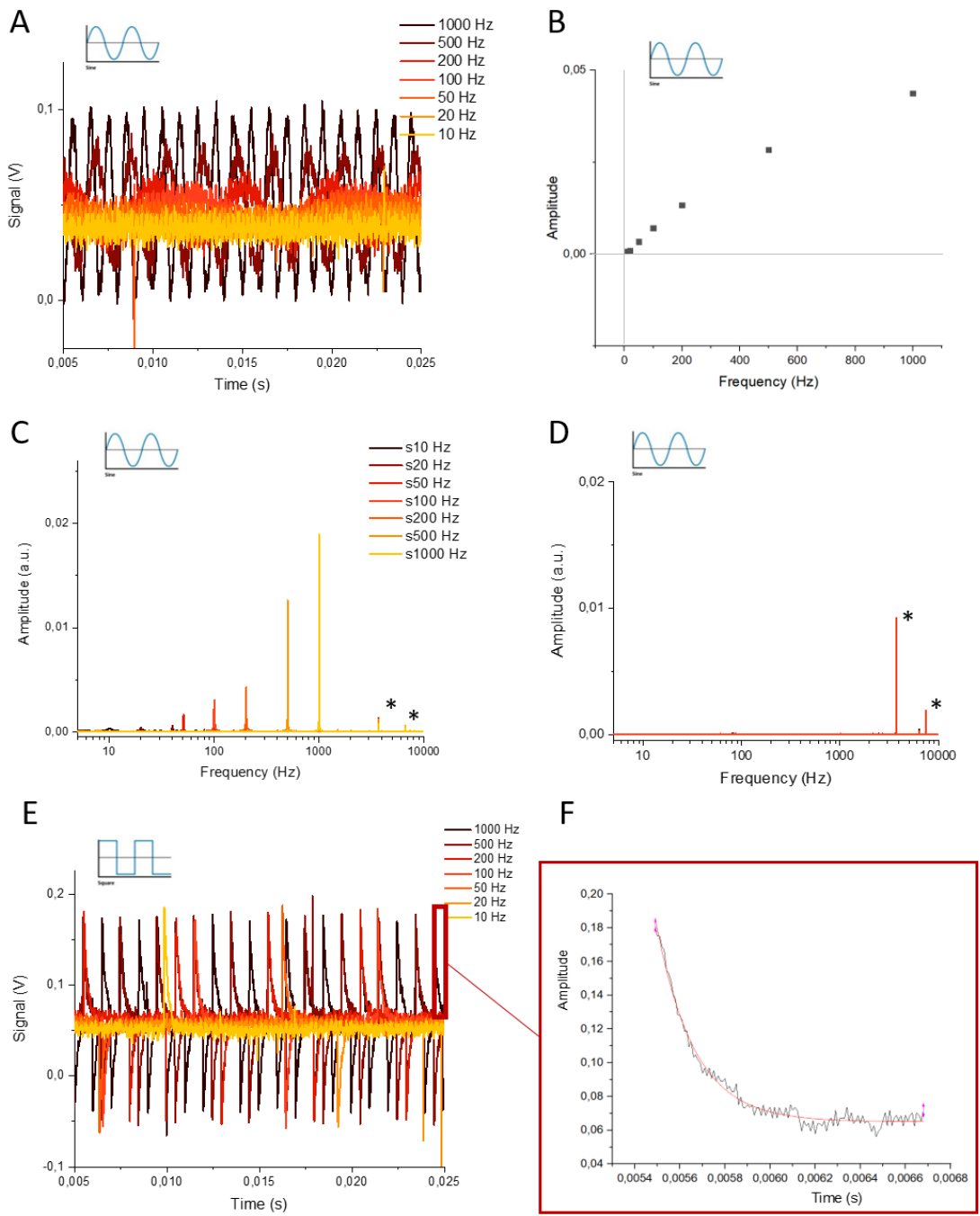

Figure 4. a) Sinusoidal response of the device when forced by a sinusoidal motion; b) Amplitude of the sinusoidal response of the device at different stimulation frequencies; c) Peak in the Fourier space related to the signal and (d) electrical noise (control) during sinusoidal motion; e) Analogic response of the device when forced by square waves; f) Example of exponential decay of the transducer signal under impulsive stimulation (black) and related fitting (red).

We tested the response of the device at several frequencies (10-1000 Hz) under both sinusoidal and a square wave stimulation. Remarkably, the device provided a reliable frequency-dependent behavior in both cases as we will discuss next.

When forced by a sinusoidal motion, the device responded with an analogue response as we show in Figure 4a. This is particularly clear from the Fourier transform (FT, Figure 4c) of the signal which resulted in single peaks exactly placed at the different oscillation frequencies. Notably, the control (i.e. shorted device) only showed a peak at about 3700 Hz that we ascribed to electrical noise (Figure 4d). In the case of a sinusoidal wave, the variation in the resistance of the device increased with the frequency of the oscillation (Figure 4b) due to the increased acceleration experienced.

To better understand the dynamics of the system, we also investigated its response under square waves of similar frequencies (Figure 4e,f). In this case, the resistance of the cantilever changed rapidly, as expected from impulsive stimulation, to then decay exponentially to its resting value with a time constant of $(1.8\pm0.1)*10^{-4}$ s. Such behavior is in agreement with that of an overdamped oscillator with a harmonic frequency in the 12.5 kHz range (as estimated for a preliminary finite element modelling simulation). Throughout all the frequencies measured, the maximum amplitude of the signal was comparable, as a result of the similar amplitude of the input signal. It is also worth noting that the characteristics of the signal did not change appreciably over time.

## 4. CONCLUSION

We have proposed a method for a rapid, two-step fabrication of micro-structured, multicomponent systems to be employed as transductors, by means of 3D DLL and metal deposition. Self-shadowing structures allowed the direct preparation of independent conductive paths of arbitrary shape. Compared to standard lithographic technique, working devices could be obtained in just two fabrication steps, saving the operators time and materials, and allowing a fast paced prototype optimization. To demonstrate the scope of this approach, we fabricated a transducer/z-axis accelerometer composed of a cantilever patterned with conductive paths which responded to external stimuli by means of a variation in the impedance. To the best of our knowledge, this is the first time a device of this kind is prepared using full-3D mask-less technique. We investigated its frequency response under sinusoidal and square waves, finding experimental agreement with an overdamped oscillator model. The electrical response of the device was reliable and reproducible over the time of the experiments, without any loss of performances. Research is currently undergoing to improve the design and the material properties of the cantilever.

## ACKNOWLEDGEMENTS

The authors acknowledge funding from the European Horizon 2020 Research and Innovation Programme under Grant Agreement No 899349 (5D NanoPrinting).